\documentclass[12pt]{scrartcl}

\pdfoutput=1
\usepackage[utf8]{inputenc} 
\usepackage{mathtools} 
\usepackage{amssymb} 

\usepackage{amsfonts}
\usepackage{amsmath}
\usepackage{makeidx}
\usepackage{epsfig}
\usepackage{epsf}

\usepackage[square, numbers]{natbib}

\usepackage{amsfonts,amsmath,amsthm,amscd} 
\usepackage[british]{babel} 

\title{Chaos synchronization: a review} 
\author{\Large{Rosário D. Laureano, Diana A. Mendes, Manuel Alberto M. Ferreira}}

\bigskip

\date {ISCTE - Instituto Universitário de Lisboa }

\usepackage{graphicx}

\begin{document}

\maketitle

%\fontencoding{T1}\selectfont Rössler

%\bigskip

%Project Title: \textbf{Research on Cohomological Dynamics and Applications in Quantum Complex Networks}

\textbf{ABSTRACT:} This article provides a self-contained comprehensive review of the phenomenon of synchronization in dynamical systems, with a particular focus on chaotic systems in both continuous-time and discrete-time contexts. 

Synchronization, initially observed by Christiaan Huygens in 1665, has evolved from the study of periodic signals to encompass chaotic systems, where the sensitive dependence on initial conditions poses unique challenges. 
This review pointed to both theoretical foundations and contributions (concepts and methods) and practical insights, reinforcing the relevance of chaos synchronization in physics, biology, engineering, social sciences, economics and communication systems. 
%The review highlights recent advances in theory and methods, offering new practical insights into the synchronization and control of chaotic systems with potential applications in physics, biology, engineering, and communications.
The study investigates various coupling schemes, such as unidirectional and bidirectional coupling, and presents stability criteria under different configurations. 

In a very concise way, some ongoing research carried out by the authors is also indicated, using Lorenz, \fontencoding{T1}\selectfont Rössler and hyperchaotic \fontencoding{T1}\selectfont Rössler systems, and quadratic maps as case studies with parameter values that lead to chaotic behaviour. Special attention is given to the stability of synchronized states and the role of multi-stability and bifurcations, and its implications to loss of synchronization. 
% This article addresses the phenomenon of synchronization in chaotic dynamical systems, focusing on both continuous-time and discrete-time frameworks. 
We highlight the role of Lyapunov exponents, eigenvalues, and Lyapunov functions in guaranteeing local and global stability of the synchronized state. 
%For discrete systems, the analysis extends to quadratic maps, where asymmetric and non-linear couplings are proposed and studied. 
%Additionally, we explore the phenomenon of multi-stability and its implications for synchronization loss. 

We aim to contribute to a broader understanding of chaos synchronization and its practical applications in diverse fields of knowledge. This text shed light on the control and stability of coupled chaotic systems, offering new perspectives on the synchronization of non-identical systems and the emergence of complex synchronization dynamics.
\vspace{5pt}


\vspace{5pt}

The first observation of synchronization was described by the Dutch physicist Christiaan Huygens in 1665 \cite{Huygens1673}. In this case, the phenomenon was evidenced by the equality of periods in mutually coupled clocks. Later, synchronization of periodic signals was detected in many other dynamical processes, which became common in other scientific areas and the subject of diverse applications in engineering. Given this original discovery by Huygens, the synchronization phenomenon is, in the classical sense of the term, associated with the ability of self-oscillatory dissipative systems to adjust their behaviours to follow a global periodic motion. It should be noted that the study of dynamical systems for a long time was based on examples of differential equations with regular solutions. If these solutions remained in a bounded region of the phase space, then they would adjust to one of two types of behaviour, a stable equilibrium point or a periodic or quasi-periodic oscillation.

Nowadays, the term "synchronization" is used in a broader sense, including chaotic systems, both discrete and continuous. Numerical analysis of chaotic populations present in nature, even heterogeneous ones, reveals that synchronization is possible and frequent but is just an example.

\section{Chaotic behaviour}

Chaos belongs to the vast field of the theory of non-linear oscillations, whose significant development began in the past century. Although in the 1920s and 1930s, Andronov proposed a classification of non-linear behaviours (1933), and Van der Pol experimentally detected noise-like oscillations in electronic circuits (1927), the experience that drove the consideration of chaotic behaviour is attributed to Lorenz \cite{Lorenz}. In 1961, working on a simplified model of atmospheric transfer with three simple non-linear differential equations, he numerically observed that making very small changes to the initial conditions had a huge effect on his solutions. This was evidence of one of the fundamental properties of chaotic dynamics, which would later be termed sensitive dependence on initial conditions. It should be noted that this property had already been investigated from a topological perspective by Poincaré and described in his monograph \textit{Science and Method} in 1903.

The sensitivity to initial conditions implies that an infinitesimal perturbation in the initial conditions of a dynamical system, in either discrete or continuous time, leads to exponential divergence of initially closed orbits over time. For many years, this property made chaos undesirable as it reduces the predictability of a chaotic system over long periods. However, the scientific community gradually became aware of a third type of behaviour in dynamics: chaotic behaviour. Some experiments, whose anomalous results had previously been explained in terms of experimental error or introduced noise, were re-evaluated and explained in terms of chaos, which then became the subject of rigorous mathematical study. A measure of sensitivity to initial conditions can be quantified by the Lyapunov exponents. However, the topological entropy and the metric entropy are quantities that, like the Lyapunov exponents, also measure the complexity of a system's dynamics.

In the mid-1970s, the term "deterministic chaos" was introduced by Li and Yorke \cite{LY1975} in a famous article titled \textit{Period three implies chaos}. This article presents the study of possible periods in periodic points of real continuous maps defined on an interval and proves that if a map has a periodic orbit of period three, then it is chaotic. However, this result can be considered a corollary of a stronger result by Sharkowski \cite{Shark} presented in a 1964 article written in Russian. This result, concerning the existence of periodic points for continuous maps, is based on a certain order of natural numbers now known as the Sharkowski order. Although there is no unique formal definition of the term "deterministic chaos", chaotic behaviour can be defined as an observable pattern that appears irregular and unpredictable over large time scales.

Many non-linear dynamical systems are dependent on control parameters in a certain parametric space. For application purposes, it is crucial to understand the evolution of the qualitative behaviour of elements of a family of dynamical systems as the parameters vary. The qualitative change in the behaviour of dynamical systems induced by the variation of one of its parameters is now known as bifurcation. The Poincaré's work — to describe the separation of equilibrium solutions in a family of differential equations — is pioneering in the importance given to the qualitative/topological changes of dynamical behaviour and in the introduction of the term "bifurcation". However, there is still no consensus on the rigorous meaning of this term. It should be noted that bifurcation theory has its roots in classical mathematics, as seen in Euler's work in the 18th century. To clarify the possible "routes to chaos", the article \cite{May1976} by May, published in 1976 in the journal \textit{Nature}, is highly relevant, as it describes the period-doubling bifurcation in the logistic map.

Geometrically, an attractor can be a single point, a curve, a manifold, or even a set with a fractal structure known as a strange attractor. The presence of a chaotic attractor in phase space, which typically has an infinite and dense set of unstable periodic orbits embedded within it, implies that it is impossible to determine the position of the system on the attractor over time, even knowing its position on this same attractor at an earlier moment. The unstable periodic orbits embedded in a chaotic attractor constitute a dynamical invariant of the system, and their systematic investigation allows for the characterization and estimation of many other dynamical invariants, such as the natural invariant measure, the spectrum of Lyapunov exponents, and the fractal dimensions. The relationship between the trajectories of the chaotic attractor and the unstable periodic orbits embedded within it has been explored in detail for hyperbolic dynamical systems, for which the separation into stable and unstable invariant subspaces is consistent (robust) under the evolution of dynamics \cite{ER1985}, and for non-hyperbolic systems with homoclinic tangencies. The infinite set of unstable periodic orbits embedded in a chaotic attractor located on some symmetric invariant manifold plays a fundamental role in the destabilization mechanism of that attractor, as it is responsible for the dynamics of phenomena such as the riddling of the basin of attraction and the bubbling of the chaotic attractor (\cite{NL1997},\cite{Lai}).

It may happen that, for different regions of the system's parametric space, different attractors (multiple attractors) exist, and as such, even the qualitative asymptotic behaviour of the system depends significantly on the initial conditions established. Even in systems with symmetry, the phenomenon of multi-stability can occur, where the coexistence of more than one attractor for a given set of parameters is observed. In this case, the qualitative asymptotic behaviour of the system cannot be predicted, as for each initial condition, it is not known a priori in which attractor the system's dynamics will eventually stabilize.

\section{Chaos synchronization}

Chaotic behaviour can be observed in natural systems, in experimental settings, and in computational models from many scientific fields, revealing itself as a robust phenomenon. The capacity of chaotic dynamics to amplify small perturbations enhances its use in achieving specific desired states in a chaotic system, without substantially altering its main dynamical properties and with a small energy expenditure. The developments in control theory and chaos synchronization theory are a consequence of this. These two processes (phenomena) have common roots. In both processes, the underlying idea is to intervene in a non-linear chaotic system by choosing parametric regions to perturb or by applying external control to restrict its movement to an accessible subspace of the phase space. In chaos control, the aim is to shift the system to an accessible invariant set where movements are more regular, while in chaos synchronization, the goal is to obtain an invariant set in the phase space of the coupled system - the synchronization subspace - where synchronous movement takes place. Although the concept of chaos synchronization has generally evolved autonomously, there is a recent trend to unify the study of chaos control and chaos synchronization under the same framework.

The dynamics of a system exhibit chaotic behaviour when it never repeats itself, and even if initial conditions are correlated by proximity, the corresponding trajectories quickly become uncorrelated. As such, the possibility of achieving synchronization of chaotic systems through appropriate coupling may seem impossible. Intuitively, we might say that the terms "chaos" and "synchronization" are mutually exclusive. Even when tracking the evolution of two identical chaotic systems that start from nearby points in phase space but are not exactly identical, we find that they diverge in their behaviours, although both retain the same chaotic attractor pattern. Two chaotic processes are observed without any mutual correlation (independent).

On the other hand, the possibility of qualitative/topological changes in the dynamical behaviour of elements in a family of dynamical systems, indicated by possible bifurcation points, complicates the synchronization of non-identical systems, even if they differ only by small parametric mismatches. An infinitesimal deviation in any of the parameters can lead to qualitatively distinct dynamics.

Moreover, two already synchronized trajectories may lose the stability of synchronization due to the influence of external noise. However, due to the ergodic property of chaotic trajectories, after a finite transient period, they will once again become close and may resynchronize. In this sense, the method of synchronization is robust against small external noise.

Studies conducted over the last three decades have shown that the intuitive idea that chaotic systems cannot behave in synchrony is false. In fact, although it is impossible to exactly reproduce identical initial conditions and equal parameters, there are sets of coupled chaotic oscillators in which the attractive effect of a sufficiently strong coupling can cancel/compensate for the tendency of trajectories to diverge due to chaotic dynamics. As a result, it is possible to achieve total synchronization in chaotic systems as long as they are coupled by an appropriate dissipative coupling. Synchronizing chaotic systems, therefore, means coupling them in such a way as to "force" them to follow the same trajectory on the chaotic attractor by applying small perturbations between the systems.

In the synchronization of two chaotic systems, it is possible to consider two coupling configurations: unidirectional (or directional) and bidirectional (mutual or global). In unidirectional coupling, one of the systems, referred to as the driver or master, influences the other, referred to as the response or slave, without the latter providing feedback to the former. Bidirectional coupling implies mutual interaction between the systems.

Coupled chaotic dynamical systems are constructed from simple, low-dimensional elements to form new, more complex organizations, ensuring that the dominant characteristics of the underlying individual components are preserved. This "cumulative" construction can also be used to create a new system whose behaviour is more flexible and/or richer than that of its components, but whose analysis and control remain accessible. The concept that several systems, when interacting non-linearly, collectively give rise to new dynamics that cannot be attributed to the individual components is known as emergence.

It is not possible to predict in advance the consequences of a coupling. The introduction of coupling in chaotic systems can drastically change the qualitative properties of dynamics. It can stabilize into periodic behaviour, it can occasionally produce hidden correlations between the elements (even when the dynamics seem turbulent), or it can induce the synchronization of a subset of the dynamical variables. The effectiveness of a coupling between systems of equal dimension depends, first of all, on the analysis of the difference between the respective coordinates of the variables of the systems involved, which can be referred to as the synchronization error. In an optimal situation, coupling chaotic systems leads to their asymptotic synchronization, where the synchronization error converges to zero over time. However, a less demanding form of synchronization was recently introduced by Stefa\'{n}ski and Kapitaniak \cite{SK2003}, where only stabilization of the synchronization error below a constant of less than one is expected. This is known as practical synchronization in the sense of Kapitaniak. It is noted that chaotic dynamics introduce new degrees of freedom into sets of coupled systems. However, when two or more chaotic oscillators are coupled and synchronization is achieved, the number of dynamical degrees of freedom for the coupled system generally decreases. Nevertheless, the specificities of chaotic behaviour make it impossible to directly apply methods developed for the synchronization of periodic oscillations to the synchronization of chaotic systems.

\subsection{Identical synchronization}

The synchronization of chaos began in the mid-1980s with the studies of Kaneko \cite{K1986} and Afraimovich \textit{et al.} \cite{AVR} on the coupling of identical discrete and continuous systems, respectively, evolving from different initial conditions. It is worth noting, in the same decade, the works of Fujisaka and Yamada (\cite{FY},\cite{FY2}), who introduced the study of transverse Lyapunov exponents in the coupled system, emphasizing how the dynamics can change with the coupling, and the works of Pikovsky (\cite{P1},\cite{P2}) and Afraimovich \textit{et al.} \cite{AVR}, who presented many of the fundamental concepts in chaos synchronization. It was only in the 1990s, with the work of Pecora and Carroll (\cite{PC},\cite{PC2},\cite{PC3}), that chaos synchronization was consolidated as an autonomous research topic, alongside the rigorous establishment of chaos control theory by Ott, Grebogy and Yorke \cite{OGY}. The aforementioned articles immediately attracted significant attention from the scientific community and triggered the development of several synchronization methods.

In communication systems, especially those involving signals whose future behaviour is difficult (or impossible) to predict during transmission, some form of synchronization between the transmitter and receiver is evidently useful. This fact motivated the work of Pecora and Carroll (\cite{PC},\cite{PC2},\cite{PC3}), where a method of synchronization is established by coupling two identical chaotic dynamical systems through transmission of a driving subsystem. This subsystem acts as a common chaotic signal between them. Thus, the method of Pecora and Carroll, also known as complete replacement, suggests how a synchronous chaotic state can be used as a driver in communication. Therefore, given a chaotic system, synchronization by this method requires decomposing the system to obtain a suitable driving subsystem. Usually, various combinations of a subset of state variables are tested to identify a stable driving subsystem. It may seem counterintuitive that a non-dissipative system could lead to synchronization, but in a multidimensional volume-preserving dynamical system, there must be at least one contracting direction so that phase-space volumes are conserved, which allows for the selection of a stable subsystem. Given the possibility of synchronizing two chaotic systems, it is necessary to determine the conditions under which such synchronization is stable. In addition to establishing the coupling mechanism, which is relatively straightforward but deceptively simple, Pecora and Carroll (\cite{PC},\cite{PC2}) also provide the first answers to this question.

Although in complete replacement there is only a finite number of possible decompositions of a chaotic system, the core idea of Pecora and Carroll to decompose chaotic systems to obtain a driving subsystem has led to other, more general methods of synchronization that do not require decomposing the original chaotic system into two stable subsystems. Kocarev and Parlitz (\cite{KP2},\cite{Par}) proposed the active-passive decomposition method by decomposing the driver system into two components: an active component and a passive one. A scalar signal from the driver system is transmitted to the response system, and this can be a function of an information signal. Different replicas of the passive component synchronize when driven by the same active component. According to Stefa\'{n}ski and Kapitaniak \cite{AK2003}, the chaotic trajectory of a system can synchronize with the chaotic trajectory of an identical system by adding a linear damping term proportional to the difference between the corresponding state variables of the two systems. This term acts as a control signal applied only to the response system in the form of negative feedback, and as such the coupling is referred to as negative feedback control. The application of this method to different experimental models by authors such as Lai and Grebogy \cite{LG}, Kapitaniak \cite{K}, John and Amritkar \cite{JA}, Anishchenko \textit{et al.} \cite{AVP}, Ding and Ott \cite{DO}, Pyragas \cite{Py2} and Wu and Chua \cite{WC} shows that it is effective when the coupled system has a single attractor.

According to Fujisaka and Yamada (\cite{FY}, \cite{FY2}), a natural way to introduce bidirectional coupling between two identical chaotic systems is to add symmetric linear coupling terms to the expressions that define them. This coupling mechanism, which can be total or partial, is called linear diffusive coupling. A study by Stefa\'{n}ski \cite{Stef} shows that the properties of exponential divergence and convergence in total coupling make it possible to estimate the largest Lyapunov exponent of any chaotic dynamical system, a possibility that is particularly useful in non-smooth systems, where the estimation of Lyapunov exponents is not straightforward.

As discussed, there are several coupling configurations between identical chaotic systems so that they become synchronized. In this regime, known as identical synchronization, the chaotic trajectories of the coupled identical chaotic systems coincide exactly over time due to the strong interaction between them. Each of the systems maintains its chaotic behaviour, but when the symmetric synchronous state is achieved - that is, when the evolution of their state vectors coincides - the dynamics of the coupled system are restricted to a synchronization hyperplane in the phase space.

In the coupling of discrete chaotic systems, parameters are considered that control the coupling strength between the systems, as in linear diffusive coupling and negative feedback control coupling of continuous systems. Depending on the structure of the coupling terms involving these parameters, several coupling schemes in discrete time are distinguished: external or internal dissipative coupling, linear diffusive coupling, quadratic coupling terms, or bilinear coupling. The stability results for the chaotic synchronous state depend on the coupling parameters considered.

\subsection{Generalized synchronization}

Although the phenomenon of chaos synchronization was detected from examples involving identical chaotic systems, the behaviour of these systems in synchrony represents only a sample of the abundance of different types of interdependent behaviour that can be detected in coupled chaotic systems. While the identical synchronization regime is the most common and has the largest number of theoretical results, many studies have revealed that it is also possible to synchronize non-identical systems, defined by evolution laws that differ only by small parametric mismatches or that are even distinct (and may even differ in dimension). In this case, the synchronization is called generalized synchronization, either in a bidirectional or even unidirectional configuration. Many of the coupling schemes between non-identical chaotic systems are extensions of those known for identical systems.

The first mathematical definition of synchronized chaos in a generalized sense was made by Afraimovich \textit{et al.} \cite{AVR} in 1986 and is based on the existence of a homeomorphism between systems with parametric mismatches that relates the projections of the synchronized chaotic trajectories onto subspaces of the phase spaces of the drive and response systems. However, the term "generalized synchronization" was only introduced by Rulkov \textit{et al.} \cite{RSTA} in 1995. Referring to the synchronization of unidirectionally coupled periodic systems, the central idea in this article, as well as in Kocarev and Parlitz's \cite{KP} article from 1996, is to take the ability to predict the current state of the response system from the chaotic information measured in the drive system as a definition of generalized synchronization. Predictability points to the existence and stability of a chaotic attractor in the coupled system.

Thus, most detection methods for generalized synchronization consider it to be represented by the existence of a functional relationship between the systems, the more regular, the better, whose graph contains an invariant and stable manifold, called the synchronization manifold, within which the chaotic attractor related to synchronization is embedded. This synchronization regime is weaker than identical synchronization, since the persistent (robust) and stable dependence between the state vectors of each system is not necessarily the identity of states. However, in general, the absence of intrinsic symmetry in the coupled system makes it difficult to obtain a stable synchronous state. It is noted that the functional relationship does not necessarily have to be valid throughout the phase space of the coupled systems but only on the invariant manifold.

It should be noted that the original definition of Afraimovich \textit{et al.} \cite{AVR}, although it allows for a set of analytical results on the stability of synchronous states, is not particularly satisfactory as it does not refer to the attractor nature of the synchronization set and requires the verification of conditions whose validity in real experiments cannot always be demonstrated. On the other hand, the definition presented by Rulkov \textit{et al.} \cite{RSTA} encompasses situations in physics, biology, and economics, in which chaos synchronization has been detected, where the requirement of a homeomorphism between the projections was not met. However, for systems with invertible dynamics, Rulkov \textit{et al.}'s definition \cite{RSTA} is equivalent to the existence of a continuous function between the states of the systems when they are on the synchronized chaotic attractor.

Studies show that the response system is asymptotically stable whenever there is a function that transforms each trajectory on the attractor of the drive system into a trajectory on the attractor of the response system, known as the synchronization function. In this case, the synchronized chaotic trajectories are located on a stable synchronization manifold. Based on the equivalence between generalized synchronization in the coupled system and the asymptotic stability of the response system, Abarbanel \textit{et al.} \cite{ARS1996} established a criterion for detecting generalized synchronization, called the auxiliary system approach.

Generalized synchronization includes identical synchronization as in a particular case, in which the functional relationship is the identity function, and the synchronization manifold is a hyperplane. However, while this is easily visualized in the representation of the difference between the coordinates of the two coupled systems, the detection of generalized synchronization does not follow a simple method, especially when analyzing information obtained experimentally. Except in special cases of coupling between systems with small parametric mismatches, it is rarely possible to provide explicit formulas for the synchronization function or to have a trivial synchronization manifold in phase space. In general, the synchronized chaotic oscillations are different from those generated by the uncoupled chaotic systems. Therefore, the analogy between the synchronized chaotic attractor and the chaotic attractors of the uncoupled systems cannot be considered a requirement to define generalized synchronization.

The asymptotic stability of the response system does not guarantee that the synchronization function is continuous, nor even the existence of a synchronization function between the systems. Experimental situations have been observed in which the response system is asymptotically stable, but the chaotic attractor of the coupled system has a complex structure, and the synchronization function is not differentiable. The dynamics on the synchronization manifold are generally quite complex, due to the lack of symmetry in the coupled system or the non-invertibility of the drive system. Contrary to what is observed in identical synchronization, where the trajectories are attracted to the symmetry plane and the synchronization manifold is trivial, many real systems exhibit synchronization subspaces with non-trivial geometric structures inherent to the coupled system: roughness, cusps, or bands, which may coexist in the same system and have a detrimental effect on the detection of synchronization. Various existing methods for detecting generalized synchronization, presented in \cite{RSTA}, \cite{PCH}, and \cite{SBJSS2002}, are hindered by the presence of such structures. 

The occurrence of roughness is generally caused by the existence of invariant sets embedded in the synchronization subspace in which the synchronization function has different degrees of Hölder regularity. The different Hölder exponents, given by the modulus of the ratio between the Lyapunov exponent relative to the transverse contracting direction and the smallest negative Lyapunov exponent of the driver, depend on the intensity of the contraction rate in the transverse direction to the synchronization subspace \cite{HOY1997}. 

The presence of cusps typically results from the existence of critical points in the attractor of the driver system defined by a smooth, non-invertible map. In the vicinity of a critical point, where the Jacobian matrix is singular, orbits of the driver system may exist along which the contraction is arbitrarily large, and the synchronization subspace is typically non-differentiable near it. 

If the driver system is non-invertible, the synchronization function may not be continuous or may not even exist as a function, as there are typical states of the driver that have more than one pre-image. In the synchronization subspace, several bands may then occur, in which case the synchronization function is generally replaced by a multivalued relationship between the coupled systems, although the response remains asymptotically stable. In this case, it is not possible to predict a response state from that of the driver.

The detection of the characteristics of generalized synchronization based on experimental information strongly relies on the continuity of the synchronization function and, in general, also requires a certain degree of smoothness. In this case, the functions in the coupled system are unknown, and when generalized synchronization is stable, an attempt can be made, in relatively simple cases, to approximate the synchronization function using numerical methods. However, if deviations from the functional dependence between the systems occur, it will never be clear whether these are due to the loss of synchronization in the coupled system or to the inaccuracy of the function considered. Thus, alternative definitions have emerged that differ in the regularity properties imposed on the synchronization function and have yielded different results in the detection of generalized synchronization in experimental information. Recently, diffeomorphic properties have been required by authors such as Abarbanel \textit{et al.} \cite{ARS1996}, Pyragas \cite{Py1996}, and Hunt \textit{et al.} \cite{HOY1997}. Pyragas \cite{Py1996} further distinguishes between two types of generalized synchronization: strong synchronization in the case of a smooth synchronization function and weak synchronization otherwise.

\subsection{Local and global stability}

For generalized synchronization to have practical (applicative) interest, it must persist under arbitrarily small perturbations, whether of the coupling or of the dynamics of the component systems. 

As in the case of identical synchronization, the stability of the synchronization manifold can be local - guaranteed by the negativity of the Lyapunov exponents that characterize the perturbations transverse to the synchronization manifold (transversal or conditional Lyapunov exponents) \cite{Py1996} and/or by the study of the eigenvalues of the linearization of the coupled system - or global, guaranteed by the existence of a Lyapunov function (Lyapunov's direct method) \cite{KP}. 

The local stability results for synchronization do not guarantee that it will hold when the coupled system is started from another initial condition. To investigate the possibility of stable synchronization in the coupled system, the choice of the initial condition is not irrelevant when there is more than one attractor in the phase space. In contrast to Lyapunov's direct method, the study of transversal Lyapunov exponents is quite straightforward and can be easily applied, even in very complex systems. However, Stefa\'{n}ski and Kapitaniak \cite{SK2003} have pointed out that, in practice, the negativity of Lyapunov exponents does not always guarantee that there are no unstable invariant sets in the synchronization manifold that could cause a loss of synchronization when noise or small parametric mismatches are present. Note that, unlike what happens in unidirectional coupling, in bidirectional coupling, the Lyapunov exponents of one of the systems are not the same as the exponents that characterize the transverse perturbations. In unidirectional coupling, the behaviour of the coupled system on the synchronization manifold is controlled solely by the dynamics of the driver system. When synchronization is lost, the driver system ceases to have full control over the behaviour of the response system, and small perturbations in the response system will grow. Although the process of loss of synchronization is similar to that observed in identical synchronization, identifying bifurcations of the bubbling or blowout type may be hindered by the complexity of the synchronization subspace. By continuously differentiating between the driver and the response as the coupling strength decreases, Barreto \textit{et al.} \cite{BSGS2000} propose a method that allows the problem to be addressed using a decomposition based on the identification of unstable periodic orbits of the driver system. The creation and evolution of a complicated set of orbits that develop outside the synchronization manifold, known as the emergent set, are described. A critical transition point for this process is also identified.

\section{Some explorations on chaos synchronization}

From this essential support, we are applying identical and generalized synchronization regimes, in various coupling schemes, to the Lorenz and \fontencoding{T1}\selectfont Rössler systems and logistic maps, using parameter values that lead to chaotic behaviour. These studies will be made available in publications that are in progress, but we can give a brief description now. 
The results obtained provide rigorous criteria for establishing local and global stability of synchronized states through methods that involve Lyapunov exponents, eigenvalues, and Lyapunov functions. This approach not only broadens our understanding of existing coupling mechanisms, but also proposes novel coupling schemes, such as partial replacements in the Lorenz and \fontencoding{T1}\selectfont Rössler systems, and innovative asymmetric couplings in quadratic discrete systems. The study of stability transitions and loss of synchronization through bubbling and blowout bifurcations further illustrates the delicate balance required to maintain synchronization in chaotic systems.

\subsection{In continuous chaotic dynamical systems}

Our research focuses on the possibility of asymptotically synchronizing pairs of continuous chaotic systems, either by unidirectional or bidirectional coupling, using Lorenz, \fontencoding{T1}\selectfont Rössler and hyperchaotic \fontencoding{T1}\selectfont Rössler systems. 

% In Section 2.1, we describe the most commonly used coupling schemes to synchronize two identical chaotic continuous dynamical systems. 
In unidirectional coupling, we explore coupling schemes by total or partial replacement, active-passive decomposition, negative feedback control, and singular value decomposition, indicating their respective stability criteria. 
% Many of the coupled systems are obtained by combining the different coupling schemes with partial replacement, particularly in the non-linear terms of the second system, or total replacement.  
We are investigating the attainment of stable synchronization when some of the coupling schemes are combined with partial replacement, particularly in the non-linear terms of the second system. As far as we know, this possibility has been little explored. Although in some schemes only local stability of the synchronous state is concluded, we explore coupling schemes where global stability is guaranteed. The global stability conditions result from the direct Lyapunov method, taking a different approach from the usual one.

We also address linear bidirectional diffusive coupling and describe, in terms of the coupling parameter, the various stability transitions of the chaotic attractor embedded in the synchronization manifold, considering the bifurcations characterized by the equations defining the coupled system. 

%Section 2.2 addresses the generalized synchronization of non-identical continuous dynamical systems. 
% In Section 2.3, we present the application of the methods described to the Lorenz (Subsection 2.3.1) and \fontencoding{T1}\selectfont Rössler (Subsection 2.3.2) systems, considering parameter values that lead to chaotic behavior in each of them. 
Some results of asymptotic synchronization stability are obtained. To investigate the local asymptotic stability, we calculate the eigenvalues of the linearized equation of the coupled system or study the transverse Lyapunov exponents. By constructing an appropriate Lyapunov function whose derivative can be bounded, we manage to exhibit conditions dependent on the coupling parameter(s) that guarantee global asymptotic stability. These conditions result from the possibility of bounding the derivative of the Lyapunov function through constants that limit the dynamical variables of the systems. 

%This section also contains a brief description of the chaotic attractors of the Lorenz and \fontencoding{T1}\selectfont Rössler systems.

\subsection{In discrete chaotic dynamical systems}

We are studying a non-linear coupling scheme that naturally arises from the family of analytical complex quadratic maps. The decomposition into real and imaginary parts of each map results in a bidirectionally coupled system of two unidimensional real quadratic maps with distinct control parameters. The resulting coupling term is proportional to the square of the difference between the dynamic variables of the component systems. To the best of our knowledge, this coupling has not been studied so far. It is an asymmetric coupling between real quadratic maps, which presents an additional challenge. 

In the study carried out, we take parameter values that lead to chaotic behaviour in the component maps. The dynamics of coupled bidimensional discrete chaotic systems are studied, obtained through asymmetric coupling. When practical synchronization is not achieved, but the difference between the variables of the two systems is bounded, a control technique was applied to the coupled system to optimize the results obtained.  The control technique used extends the well-known Ott-Grebogy-Yorke chaos control method \cite{OGY}, through a small perturbation of the coupling parameter. 

In addition, we explore some variants of the original coupling. We obtain stable identical and generalized synchronization when we consider some variants of the original coupling, favouring the absence of symmetry in the coupling. Two of these constitute a generalization to the use of distinct coupling parameters.
%We consider the particular case of coupling between identical maps. 
We continue to consider a single coupling parameter as we aim to investigate the advantages of symmetry in the coupled system. This is still a bidirectional coupling between identical systems that, to our knowledge, has also not been studied. 
We also explore the study of a generalization of the coupling to the case of distinct coupling parameters in each of the component systems: the case of identical and non-identical maps. 

% Section 3.4 consists of the study of the same coupling, but in its respective unidirectional configuration. This is an asymmetric coupling, although we consider identical maps. 
For all couplings, we analyze the dynamics of the coupled system and, subsequently, investigate the local asymptotic stability of synchronization, based on the eigenvalues of the linearized equation of the transverse system and the estimation of the corresponding Lyapunov exponents. By analyzing the difference between the dynamical variables of the systems, we establish some results that guarantee their stable synchronization. The computational approaches presented, in addition to confirming these results, shed light on the necessary study in each coupling scheme to determine the parameter values that lead to stable synchronization.

\section{Conclusion}

This review study provides a self-contained approach to the preliminary, but fundamental, concepts and results in synchronization of dynamical systems, with emphasis in chaotic dynamical systems, and establishes the terminology necessary for the research work carried out. By tracing the evolution from classical concepts to modern generalizations, this article reveals how the phenomenon of synchronization extends beyond periodic oscillations to the realm of chaotic behaviour with profound theoretical and practical implications. The identification and characterization of identical and generalized synchronization regimes have demonstrated the richness and complexity of the behaviour of coupled systems. We have provided an overview of the key types of synchronization, identical and generalized synchronization, and discussed their theoretical underpinnings. The stability analysis of coupled systems was presented using tools such as Lyapunov exponents, eigenvalues, and Lyapunov functions, allowing local and global stability criteria to be established. We highlighted the challenges posed by multi-stability, bifurcations, and the loss of synchronization through bubbling and blowout transitions. Special attention was paid to the role of chaotic attractors, which serve as a foundation for the control and predictability of synchronized states.

There are several reasons that motivated the choice of this research topic. The phenomenon of chaos synchronization is, from the outset, interesting due to its high application potential. This potential is transversal to knowledge areas as distinct as physics, biology, engineering, or economics. We find it particularly stimulating to study a phenomenon that requires the adjustment of dynamical behaviours to obtain coincident chaotic motion, which is possible precisely in chaotic dynamical systems where sensitive dependence on initial conditions is one of the characteristics. The existence of certain analogies between synchronization and chaos control and the possibility of applying chaos control techniques as a way to optimize the results of synchronization constitute yet another motivating factor for choosing the topic. Moreover, we do not overlook the fact that the phenomenon of chaos synchronization is quite recent in the non-linear theory of dynamical systems and continues to generate significant interest in the scientific community. However, despite the efforts made to investigate this phenomenon, many questions remain open.

From secure communication systems to control applications in biological and engineering networks, synchronization has the potential to become a powerful tool in system design. The insights into multi-stability, attractor co-existence, and the role of chaotic trajectories in synchronization loss provide fertile ground for future research, particularly in applications where robustness against parameter mismatches and noise is crucial. The diversity of approaches, including analytical techniques, computational experiments, and theoretical generalizations, demonstrates that synchronization is a versatile and evolving research area. By bridging the gap between classical synchronization theory and modern chaos theory, this work contributes to a broader understanding of complex systems. It opens avenues for new studies on the dynamics of non-identical systems, the interplay between chaos and control, and the effects of coupling asymmetries on global stability. The proposed techniques and methodologies can inspire further exploration, promoting the design of more efficient coupling protocols, and enhancing the predictability and stability of synchronized states in real-world applications.

This review serves as a resource for researchers and practitioners interested in understanding and harnessing synchronization in diverse applications, from secure communications to the control of biological and engineering networks. The insights presented pave the way for future investigations into the synchronization of non-identical systems, the impact of asymmetries in coupling, and the development of robust synchronization protocols that resist parameter mismatches and noise. The concepts discussed here have significant potential for real-world applications, underscoring the continued relevance of synchronization in both the theoretical and applied sciences.

% We present a brief review of preliminary notions in non-linear dynamics and, in particular, in the synchronization of chaotic dynamical systems. 
% By allowing us to address the essential questions regarding the synchronization phenomenon, all the couplings studied involve only two chaotic systems.

\bigskip

%Chapter 1 aims to introduce the fundamental concepts and results for the study of chaos synchronization and to establish the terminology necessary for the research work carried out. Section 1.1 consists of a brief review of general elements in non-linear dynamics, both in continuous and discrete time. In Section 1.2, we present a brief introduction to the classical theory of synchronization, exposing the main synchronization regimes between identical or non-identical chaotic dynamical systems. For the regimes of identical and generalized synchronization, we have chosen to address the stability of the synchronous state, with the determination of synchronization thresholds, and the transition/loss of synchronization from a topological point of view only in external dissipative coupling and generalized external dissipative coupling, respectively. This choice is due to the fact that these coupling schemes are the most realistic from an applicative point of view, particularly in the modelling of interconnected populations. Although it is not one of the regimes chosen for our study, we briefly present phase synchronization. We have chosen to address in this section only the elements related to the synchronization of discrete chaotic systems (which best models natural couplings between populations), as the methods of synchronization in continuous systems are presented and applied in Chapter 2.

\bigskip

%This concludes summarizing the content and structure of the thesis, and outlining the main goals and methodologies employed in the study of the synchronization of chaotic systems. Each chapter builds on the preceding ones, establishing a comprehensive and detailed investigation of the synchronization phenomenon in both continuous and discrete time systems. 

%CONCLUSION 1:
%The theoretical contributions presented here are of direct practical relevance. 

\vspace{5pt}

%CONCLUSION 2:
% The review underscores the importance of coupling schemes, with a focus on continuous and discrete systems, as well as innovative coupling techniques that promote stable synchronization. 

\end{document}